\input amstex
\magnification=\magstep1
\documentstyle{amsppt}

\NoBlackBoxes

%
%
\newcount\equanumber \equanumber=0
\def\eqlab #1%
    {%
    \LookUp Eq_#1 \using\equanumber%
    {\label}%
    }%
%
%
\newcount\eqauto    \eqauto=0
\def\eqnum{\global\advance\eqauto by 1
                  \eqlab{{\jobname.\the\eqauto}} }
%
%
\newcount\thmnumber \thmnumber=0
\def\thmlab #1%
    {%
    \LookUp Eq_#1 \using\thmnumber%
    {\label}%
    }%
%
%
%
\newcount\thmo    \thmo=0
\def\thmnum{\global\advance\thmo by 1
                  \thmlab{{\jobname.\the\thmo}} }
%
%
\newcount\refnumber \refnumber=0
\def\reflab #1%
    {%
    \LookUp Ref_#1 \using\refnumber%
    {\label}%
    }%
%
%
\def\LookUp #1 \using#2%
    {%
    \expandafter\ifx\csname#1\endcsname\relax
        \global\advance #2 by 1
        \expandafter\xdef\csname#1\endcsname{\number#2}%
    \fi
    \xdef\label{\csname#1\endcsname}%
    }%
\TagsOnRight
\font\script=rsfs12 at 10pt

\def\ie{{\it i.e.}}
\def\etc{{\it etc}}
\def\p{\partial}
\def\d{\text{d}}
\def\Q{\Cal Q}
\def\D{\Cal D}
\def\hor{\goth{hor\,}}
\def\a{\alpha}
\def\b{\beta}
\def\fb{\bar f}

\def\om{\Omega}
\def\Ham{\text{\script H}}
\def\Op{\text{\script O}}
\def\L{\text{\script L}}

\def\V{\Cal V}
\def\P{\Cal P}
\def\Pb{\overline p}
\def\Qb{\overline q}
\def\Or{{\Cal O}}
\def\aYM{\alpha_{_{\text{YM}}}}

\def\R{\bold R}

\def\Chi{\bold X}
\def\<{\langle}
\def\>{\rangle}
\topmatter
\title 
Geometric quantization on symplectic fiber bundles
\endtitle
\author Yihren Wu\endauthor
\affil Department of Mathematics, Hofstra University, Hempstead, NY 11550
\endaffil
\endtopmatter
\centerline{\bf Abstract}

Consider a fiber bundle
in which the total space, the base space and the fiber are all symplectic
manifolds. We study the relations between the quantization of these 
spaces. In particular, we discuss the geometric
quantization of a vector bundle, as oppose to a line bundle, over
the base space that recovers the standard geometric quantization of the
total space.

\vfil\pagebreak
\subheading{I. Introduction}

In \cite{\reflab{GLSW}}, 
the authors considered a family of symplectic manifolds
and gave a topological condition under which these manifolds can be
quantized simultaneously using the procedures of geometric quantization
\cite{\reflab{Sn}} \cite{\reflab{woodhouse}}. 
More precisely, let $E\to B$ be a differentiable fiber bundle
each fiber $E_b$ is equipped with a symplectic form $\om_b$ so that
$\{E_b,\om_b\}$ is symplectomorphic to the model 
fiber $\{F,\om_F\}$, with  structure group of this bundle
preserving the symplectic form. A closed two-form $\om_E$ on $E$ is termed
``closed extension'' of the family of symplectic
forms $\om_b$ if $\om_E=\om_b$ when restricted to the fiber.
The authors showed that
closed extension exists if it exists at the cohomology level. 
In which case, the line bundle $\L\to E$ whose curvature is $\om_E$ pulls
back to a prequantization line bundle $\L_b\to E_b$. This then allows one
to carry out the necessary calculations on $\L$ as a way to quantize
the whole family of symplectic manifolds $\{E_b\}_{b\in B}$.

We are interested in the situation where the total space
$E$, the base space $B$ and the fiber $F$ are all symplectic manifolds, and
we want to study the relation between the quantization of these spaces.
It is known that if the bundle $E\to B$ is trivial, then the relation between
the quantum Hilbert spaces is given by the tensor product:
$\Q(E)=\Q(F)\otimes\Q(B)$, where $\Q(X)$ denotes the quantum Hilbert
space of the classical phase space $X$, suppressing in our notation the 
dependence on the polarizations chosen for the quantization \etc.

Suppose there is a polarization on $F$ that is preserved by the structure
group of the fiber bundle, then there is an associated vector bundle
$\Cal V\to B$ whose fiber is $\Q(F)$.
Our  goal is to give conditons and procedures for
quantizing this ``prequantization vector
bundle'' that parallels the standard geometric quantization on line bundles.
The resulting wavefunctions, \ie, sections on this vector bundle 
covariant constant along a polarization on $B$, are vector-valued
wavefunctions. This will provide a nice setup for multicomponent
WKB method, since the quantization of $E$ has both a scalar
version ($\L\to E$) and a vector version ($\V\to B$).
Multicomponent WKB theory has gain recent interest starting with the work
of \cite{\reflab{littlejohn}}.  

A detail setup for the quantization of vector bundles will be discussed in
section 2, where we state our main theorem. In section 3 we show that
the case of a particle moving under the influence of an external Yang-Mills
field fits into our fiber bundle formality. This represents a non-trivial
example where the whole system can be quantized, and the resulting
Hilbert space is a twisted tensor product induced by the Yang-Mills 
potential. The prove of the theorem will occupy section 4.

\subheading{II. The Setup}
\vskip.4cm
Let $\{E,\om_E\}$, $\{B, \om_B\}$ and $\{F,\om_F\}$ be symplectic manifolds,
we assume there are canonical one--forms $\a_B$, $\a_E$, $\a_F$ such that
$\d\a_B=\om_B$, $\d\a_E=\om_E$ and  $\d\a_F=\om_F$.
Let $\P(F)$ be a polarization on $F$. 
Let $\pi:E\to B$ be a symplectic fiber bundle with fiber $F$, which we 
assume to be compact and simply connected.
We further
assume the structure group preserves the canonical one--form and the
polarization. More precisely, there exist a local trivilization
$\chi_{_i}:U_i\times F\to E$ such that firstly, if 
$b\in U_i$, then $\chi_{_{i,b}}=\chi_{_i}(b,-):F\to E_b$ satisfies
$\chi_{_{i,b}}^*(\a_E|E_b)=\a_F$. So $\om_E$ is a closed extension
of the symplectic forms on the fibers $E_b$ in the sense of 
\cite{\reflab{GLSW}}.
Secondly, if
$b\in U_i\cap U_j$, then
$\chi_{_{ij,b}}=\chi_{_{j,b}}^{-1}\chi_{_{i,b}}:F\to F$ preserves the
polarization
$\chi_{_{ij,b*}}\P(F)=\P(F)$. 

As a closed extension, the symplectic form $\om_E$ defines an Ehresmann
connection on the bundle $E\to B$ so that for $e=\chi_{_i}(b,f)$,
 $u\in T_eE$ is horizontal if
$\om_E(u,\chi_{_{i,b*}}\xi)=0$ for all $\xi\in T_fF$. 
 Note that  the Ehresmann connection
is defined up to a two--form on $B$.
We denote by
$v^\#\in T_eE$ the horizontal lift of $v\in T_bB$,
and by $\hor(B)$ the set of horizontal vector fields. We make the following
observation:
\proclaim{Proposition \thmlab{horvector}}
Let $\a_\nabla=\a_E-\pi^*\a_B$, and
let $\a^i=\chi^*_{_{i}}\a_\nabla$ be the one--form on $U_i\times F$, $v$ a 
vector field on $U_i$, then 
$$v^\#=v-\Ham_{\<\a^i,v\>}\tag{\eqlab{horvec}}$$ 
where we treat $w=\<\a^i,v\>$ as a function on $F$ with $b$ fixed, and 
$\Ham_w$ is the Hamiltonian vector field on $F$ defined through the equation
$\om_F(\Ham_w, -)=-\d_Fw$.\endproclaim
\demo{Proof}
We must show that $\chi^*_{_{i}}\om_E(v-\Ham_{\<\a^i,v\>},\xi)=0$ 
for all vector fields $v$ on $B$ and $\xi$ on $F$.
Note that
$$\om_F(\Ham_w,\xi)=-\xi(w)=\Ham_w\<\a_F,\xi\>-\xi\<\a_F,\Ham_w\>-
\<\a_F,[\Ham_w,\xi]\>.\tag{\eqlab{dF}}$$
Using the facts that $[v,\xi]=0$ and $v\<\a^i,\xi\>=v\<\a_F,\xi\>=0$, 
we compute
$$\align
\d\a^i(v-\Ham_w, \xi)&=(v-\Ham_w)\<\a^i,\xi\>-
\xi\<\a^i,v-\Ham_w\>-\<\a^i,[v-\Ham_w,\xi]\>\\
&=v\<\a_F,\xi\>-\Ham_w\<\a_F,\xi\>-\xi\<\a^i,v\>+\xi\<\a_F,\Ham_w\>
+\<\a_F,[\Ham_w,\xi]\>\\
&=-\xi\<\a^i,v\>+\xi(w)=0,\qquad \text{since $w=\<\a^i,v\>$.}\qed
\endalign$$
\enddemo

Suppose we have a
polarization $\P(E)$ on $E$ that is composed of $\P(F)$ and horizontal vector
fields. More precisely, let $e=\chi_i(b,f)$, $u\in\P_e(E)$ implies there
is a $u_F\in \P_f(F)$ and a $u_B\in T_bB\otimes\bold C$ so that
$u=\chi_{_{i,b*}}u_F+u_B^\#$. So by abuse of notation we consider $\P(F)$ 
also as
a foliation on $E$ via the trivializing maps $\chi_{_i}$, our assumption
that $\chi_{_{ij}}$ preserves $\P(F)$ implies this push forward is independent
on the coordinate patch we choose. And we denote the polarization
$\P(E)=\D^\#(B)\oplus\P(F)$, 
where $\D(B)$ is a distribution on $B$ and
$\D^\#(B)$ its horizontal lift. 

Let $\L_F$ be a prequantization line bundle on $F$ and denote by $\Q(F)$ the
space of sections on $\L_F$ covariant constant along the polarization $\P(F)$,
and we will treat $\Q(F)$ as the quantum Hilbert space for the quantization
of $F$, ignoring potential complications that may arise from the half-form
bundle that appears the process of geometric quantization. Since $F$ is 
assumed compact, $\Q(F)$ is finite dimensional, we fix a basis 
$\{\phi_1,\dots,\phi_n\}$. We denote the quantization of the 
symplectomorphisms $\chi_{_{ij,b}}$ by $\Chi_{ij,b}:\Q(F)\to\Q(F)$.
Let $\V\to B$ be the unitary vector bundle whose transition functions 
given by $\Chi_{ij,b}$ and denote by $\Gamma(\V)$ the space of sections.

Let $\L$ be a prequantization line bundle over $E$ which pulls back
to $\L_F$ by $\chi_{_{i,b}}$. That is, there is a bundle map
so that the diagram
$$\CD
\L_F@>>>\L_b\\
@VVV                @VVV\\
F@>>{\chi_{_{i,b}}}>E\endCD$$
commutes. This will be the case if the connection one form $\a_E$
on $E$  pulls back to the connection form $\a_F$ on $F$ via $\chi_{_{i,b}}$,
as we have assumed.
Since by assumption the transition functions $\chi_{_{ij,b}}$ preserve both
the connection one--form and the polarization, their quantization is 
by substitution:
$$\Chi_{_{ij,b}}\phi(f)=\phi(\chi_{_{ij,b}}(f)).\tag{\eqlab{subs}}$$

Denote by $\Gamma_{\P(E)}(\L)$ the
space of sections on $\L$ covariant constant along the polarization
$\P(E)$,
and by $\Gamma_{\P(F)}(\L)$ the space of
sections  covariant
constant along the foliation $\P(F)$. Let $\psi\in \Gamma_{\P(F)}(\L)$, 
when restricted to a local trivilization patch
$U_i\times F$, $\psi(\chi_{_i}(b,f))$ must be of the form 
$\Psi^i_\mu(b)\phi_\mu(f)$. (We will frequently suppress the index $i$
if no confusion will arise.) Thus we have established an isomorphism
$\Gamma_{\P(F)}(\L)\to\Gamma(\V)$. Consider the composition
$$\Gamma_{\P(E)}(\L)\to\Gamma_{\P(F)}(\L)\to\Gamma(\V),\tag{\eqlab{setup}}$$
the image of $\Gamma_{\P(E)}(\L)$ defines a subspace 
$\Gamma_c(\V)$ in $\Gamma(\V)$ that plays the role of covariant
constant sections. One would like to have a connection on the vector bundle
$\V$ in order to define this notion of covariant constant section on $\V$. 
For technical reasons, we need to assume that 
$$[\hor(B),\P(F)]\subseteq\P(F).\tag{\eqlab{Assume}}$$
where $[,]$ denotes vector field commutator.

Let $H:F\to\R$, we denote the prequantization operator \cite{\reflab{Sn}}
$$\Op(H)=-i\Ham_H-\<\a_F,\Ham_H\>+H.\tag{\eqlab{prequan}}$$
Through out this paper we will let $\Ham_H$ to denote the Hamiltonian vector 
field on $F$ with respect to 
the symplectic form $\om_F$, and the $b$ dependence that $H$ may have
will be treated as constant in this regard. 
And $\<,\>$ is the pairing between 
differential forms and vector fields, 
the inner product on $\Q(F)$ is denoted by $\ll | \gg$.
Let $U$ be the unitary group of the (finite-dimensional) Hilbert space
$\Q(F)$ and $\goth u$ its Lie algebra, define 
$$\align A^i(b):T_bB&\to\goth u\tag{\eqlab{connection}}\\
\left[A^i(b)v\right]_{\mu\nu}
&=i\ll \phi_\nu\,|\,\Op\left(\<\chi^*_{_{i}}\a_\nabla,v\>\right)\phi_\mu\gg
\endalign$$
for all $b\in U_i$.
We now summarize our assumptions and state our main theorem:
\proclaim{Theorem} 
Let $E\to B$ be a symplectic fiber bundle with fiber $F$ and
transition functions
$\chi_{_{ij,b}}:F\to F$ preserving the connection one--form $\a_F$ and
polarization $\P(F)$. Let $\om_E=\d\a_E$ be the symplectic form on $E$
that is a closed extension of $\om_F$ on the fibers. Suppose the Ehresmann
connection defined by $\om_E$ satisfies (\eqlab{Assume}).  Then
$A^i$ in (\eqlab{connection}) defines a connection on the vector bundle
$\V\to B$.  Let $\P(E)=\D^\#(B)\oplus \P(F)$ be a
polarization on $E$, and $\Gamma_c(\V)$ the subspace of sections
covariant constant along $\P(E)$. Then $\Psi\in \Gamma_c(\V)$ if
$v\Psi=i\<\a_B,v\>\Psi+\Psi A(v)$ for all $v\in \D(B)$.
\endproclaim
In this manner, the covariant constant condition is given in terms
of the canonical one--form on $B$, together with   
the connection $A$.  In general it is not true that
a polarization $\P(B)$ on $(B,\om_B)$ will result in a polarization
$\P^\#(B)\oplus \P(F)$ on $E$. In vector bundle quantization, one has
to deal with the given symplectic form on $B$, and the hidden symplectic
form on the total space $E$ which induce a connection $A$ on the bundle.
In multi-component WKB theory one faces a similar 
situation with two symplectic forms \cite{\reflab{littlejohn}}.

In our formulation we have imposed two rather strong technical conditions;
that the transition functions preserve the canonical form $\a_F$ and the 
polarization $\P(F)$, and that the commutator between the horizontal 
vector fields and $\P(F)$ remains in $\P(F)$. 

The conditions on transition functions are not entirely necessary, it
allows us to keep the technical details to a minimum, in that the
quantization of the symplectic transforms $\chi_{_{ij,b}}$ is given
by (\eqlab{subs}). Quantization of non-polarization 
preserving symplectomorphisms
can be obtained through the use of BKS-pairing \cite{\reflab{BKS}}.   

Condition (\eqlab{Assume}) can be considered as a 
``minimal coupling'' condition. In the standard Dirac quantization, 
the momentum variable is quantized as
$p\mapsto \tfrac{\d\ }{\d q}+A$ where $A$ is a vector potential.
This assignment presumes an interplay between the polarization (in this
case the vertical polarization so that the wavefunctions are functions
on the configuration space) and the connection $A$. Our assumption
(\eqlab{Assume}) represents an interrelation of this kind.

\subheading{III. Particles in an external Yang-Mills field}

The case of Yang-Mills field can be described as follows 
\cite{\reflab{Robson}}; Let $N\to Q$ be a principal bundle with group $G$
and connection $\aYM$, the Yang-Mills potential, 
there is an $\aYM-$dependent projection $T^*N\to T^*Q$.
The $G-$action on $N$ can be lifted to a Hamiltonian $G-$action
on $T^*N$ with moment map $J:T^*N\to \goth g^*$. Let 
$\mu\in \goth g^*$ such that the coadjoint orbit $\Or_\mu$ is integral,
and 
$E=J^{-1}(\Or_\mu)/G\sim J^{-1}(\mu)/H$ be
the Marsden-Weinstein reduced space \cite{\reflab{MW}}, where
$H$ its isotropy subgroup of $\mu$. Let $B=T^*Q$, then $E\to B$ is
a fiber bundle whose fiber is 
$F=\Or_\mu$, the transition function is induced by $G-$action
on $F$, which preserves the canonical one--form on $\Or_\mu$. 
Moreover, there is a standard polarization, the
positive K\"ahler polarization, which is $G-$invariant. 
Quantization of
$F$ with respect to this polarization gives a irreducible representation
space $\Q(F)$ of $G$ induced by the $U(1)$ representation of $H$ 
\cite{\reflab{Ki}}.
These are interpreted physically as the internal symmetries of a particle
of ``charge'' $\mu$ in the configuration space $Q$. The vertical polarization 
on $T^*Q$ lifts to a distribution on $E$ satisfying condition 
(\eqlab{Assume}). Thus all the assumptions we laid out in the previous
section hold.

Denote by $N^\#$ the principal $G-$bundle over $T^*Q$ by the pullback
$$\CD N^\# @>>> N\\
@VVV     @VVV\\
T^*Q @>>> Q\endCD$$
There is a diffeomorphism $N^\#\to J^{-1}(\mu)\subset T^*N$ given by
$$\xi=\pi^*(n)p+\mu\cdot\aYM(n)\tag{\eqlab{nsharp}}$$
where $\xi\in T^*_nN$, $p\in T^*_{\pi(n)}Q$,
$\aYM(n)\in T^*_nN\otimes\goth g$ and the dot product refers
to the pairing between $\goth g^*$ and $\goth g$. With these, the 
prequantization vector bundle on $T^*Q$ can be conveniently described
as 
$$\V=N^\#\times_G\Q(F)\to T^*Q.$$

Fix $b=(p,q)\in B$, and $v\in T_bB$,
there is a natural projection $\Pi:T_bB\to T_qQ$. Let $\chi_{_i}:Q\to N$ be
a local coordinate patch. In this setting, $\a_E-\pi^*\a_B=\mu\cdot\aYM$
and the function $w=\<\chi^*_{_{i}}\a_\nabla,v\>:\Or_\mu\to\bold R$ becomes
$$w(gH)=\mu\cdot Ad_{g^{-1}}\<\chi^*_{_{i}}\aYM(\chi_{_{i}}),\Pi(v)\>,$$
where we have identified $\Or_\mu=G/H$.
Let $\goth u$ be the Lie-algebra of unitary 
transforms on $\Q(F)$ and $\rho$ the Lie-algebra representation
$\rho:\goth g\to \goth u$. Then
one can show that the connection $A$ defined in (\eqlab{connection}) is
given by 
$$A^i(b)(v)=\rho\<\chi^*_{_{i}}\aYM(\chi_{_{i}}),\Pi(v)\>
\tag{\eqlab{yangmill}}$$
The Yang-Mills potential $\aYM$ also plays a cruicial role 
in the quantization of observables
on $T^*Q$. In particular, the kinetic energy of the particle
is quantized to $\tfrac{1}{2}(-\Delta_\a+R/6)$ where
$\Delta_\a$ is the covariant Laplace-Beltrami operator 
with respect to $\aYM$, and $R$ is the Ricci scalar 
curvature of the Riemannian manifold $Q$ \cite{\reflab{Wu}}.

\subheading{IV. Proof of theorem}

Our assertion will be proved with a sequence of propositions.
We must first show that the inner product in (\eqlab{connection}) is 
well defined, recall $\a^i=\chi^*_{_{i}}\a_\nabla$,
\proclaim{Propostion \thmlab{welldefined}}
If $\phi\in\L(F)$ is covariant constant along $\P(F)$,
then $\Op(\<\a^i,v\>)\phi$ is also covariant constant along
$\P(F)$. Thus $\Op(\<\a^i,v\>):\Q(F)\to\Q(F)$.\endproclaim
\noindent
Once we have established that $A^i$ is well defined, we must then show that 
$A^i$ transforms like a gauge:
\proclaim{Proposition \thmlab{gauge}} If $b\in U_i\cap U_j$ then
$A^j(b)=\Chi_{ij,b}A^i(b)\Chi^{-1}_{ij,b}
+d\Chi_{_{ij,b}}\Chi^{-1}_{_{ij,b}}$.\endproclaim
\noindent
Lastly we will show that the connection $A$ has the desired property:
\proclaim{Propostion \thmlab{covariant}} Let $\Psi\in\Gamma(\V)$ and
$\psi=\Psi_\nu\phi_\nu$ be the corresponding section in $\L$, let
$v\in D(B)$. If $v^\#\psi=i\<\a_E,v^\#\>\psi$, then
$v(\Psi)=\Psi\left(i\<\a_B,v\>I+A(v)\right)$, where $I$ is the
identity matrix so that $i\<\a_B,v\>I\in \goth u$.\endproclaim
\demo{Proof of Proposition \thmlab{welldefined}}
Let $\phi\in\L(F)$ covariant constant along $\xi\in\P(F)$, then
$$\xi\phi=i\<\a_F,\xi\>\phi.\tag{\eqlab{given}}$$
For any fixed $b\in U_i$ consider the operator 
$\Op(w)=-i\Ham_w-
\<\a_F,\Ham_w\>+w$ where $w=\<\a^i,v\>$ is treated as
a function on $F$. Then
$$\align
\quad\xi\Op(w)\phi-i\<\a_F,\xi\>\Op(w)\phi
=&\left(w-\<\a_F,\Ham_w\>\right)
(\xi\phi-i\<\a_F,\xi\>\phi) \tag{\eqlab{begin}}\\
&-i\xi\left(\Ham_w\phi\right)-
\left(\xi\<\a_F,\Ham_w\>\right)\phi
-\<\a_f,\xi\>\Ham_w\phi+\left(\xi w\right)\phi
\endalign$$
Here the right hand side of first line in (\eqlab{begin}) vanishes because
of (\eqlab{given}). Using (\eqlab{dF})
we have
$$\xi\<\a_F,\Ham_w\>\phi=(\Ham_w\<\a_F,\xi\>)\phi-
\<\a_F,[\Ham_w,\xi]\>\phi+(\xi w)\phi.$$
Equation (\eqlab{begin}) then becomes
$$\align
&-i\xi(\Ham_w\phi)-\Ham_w(\<\a_F,\xi\>)\,\phi+
\<\a_F,[\Ham_w,\xi]\>\phi-\<\a_F,\xi\>\Ham_w\phi\\
=\ &-i\xi(\Ham_w\phi)-\Ham_w(\<\a_F,\xi\>\phi)+\<\a_F,[\Ham_w,\xi]\>\phi\\
=\ &-i\xi(\Ham_w\phi)+i\Ham_w(\xi\phi)+\<\a_F,[\Ham_w,\xi]\>\phi\\
=\ &i\left([\Ham_w,\xi]\phi-i\<\a_F,[\Ham_w,\xi]\>\phi\right)\\
=\ &0\endalign$$
since $v-\Ham_w$ is horizontal by proposition \thmlab{horvector},
and assumption (\eqlab{Assume}) implies
$[v-\Ham_w,\xi]=-[\Ham_w,\xi]\in\P(F)$.\qed
\enddemo
\demo{Proof of proposition \thmlab{gauge}}
Here we show $A$ defines a connection on the vector bundle $\V$. 
For $b\in U_i\cap U_j$ and $v\in T_bB$, we let 
$[A^i(b)v]_{\mu\nu}= i\ll\phi_\nu\,|\,\Op(w^i)\phi_\mu\gg$ and
$[A^j(b)v]_{\mu\nu}= i\ll\phi_\nu\,|\,\Op(w^j)\phi_\mu\gg$ where
$w^i(b,f)=\<\a^i(b,f),v\>$, 
$w^j(b,\fb)=\<\a^j(b,\fb),v\>$
with $\chi_{_{i}}(b,f)=\chi_{_{j}}(b,\fb)=e$, so $\fb=\chi_{_{ij,b}}f$,
and $\a^i(b,f)=\chi^*_{_{i}}\a_\nabla(e)$, 
$\a^j(b,\fb)=\chi^*_{_{j}}\a_\nabla(e)$.
We calculate
$$\align
w^i(b,f)&=\<\chi^*_{_{i}}\a_\nabla(e),v\>=\<\chi^*_{_{ij}}\a^j(b,\fb),v\> 
=\<\a^j(b,\fb),\chi_{_{ij,*}}v\>\\
&=\<\a^j(b,\fb), v+J_*v\>=w^j(b,\fb)+\<\a^j(b,\fb),J_*v\>\\
&=w^j(b,\fb)+\<\a_F(\fb),J_*v\>
\tag{\eqlab{wi}}\endalign$$
where $J_*$ is the block matrix in
$$\chi_{_{ij*}}=\bmatrix I&0\\J_*&\chi_{_{ij,b*}}
\endbmatrix:T_bB\times T_fF
\to T_bB\times T_{\fb},F$$
so that $J_*v\in T_{\fb}F$ is a vertical vector.
Denote the last term in (\eqlab{wi}) by $\b(b,f)$, then 
we have $w^j(b, \fb)=w^i(b,f)-\b(b,f)$.
Writing out explicitly the
dependence on $F$
$$\align
[A^j(b)v]_{\mu\nu}&=i\ll\phi_\nu(\fb)\,|\,\Op(w^j(b,\fb)\phi_\mu(\fb)\gg\\
&=i\ll\phi_\nu(\fb)\,|\,\Op(w^i(b,f)\phi_\mu(\fb)\gg
   -\,i\ll\phi_\nu(\fb)\,|\,\Op(\b(b,f))\phi_\mu(\fb)\gg
\tag{\eqlab{split}}\endalign$$

The first term in (\eqlab{split}) reduces to
$$\align i\ll\phi_\nu(\fb)\,|\,\Op(w^i(b,f)\phi_\mu(\fb)\gg&=
i\ll\Chi_{\nu\tau}\phi_\tau(f)\,|\,\Op(w^i(b,f)
\Chi_{\mu\sigma}\phi_\sigma(f)\gg\\
&=\overline{\Chi}_{\nu\tau}\Chi_{\mu\sigma}[A^i(b)v]_{\sigma\tau}
=\Chi_{\mu\sigma}[A^i(b)v]_{\sigma\tau}\overline{\Chi}^t_{\tau\nu}
\tag{\eqlab{firstterm}}
\endalign$$
which gives the familiar adjoint action $\Chi A^i \Chi^{-1}$ in
matrix notation, here $\Chi=\Chi_{ij,b}$ as in (\eqlab{subs}).

As for the second term in (\eqlab{split}), suppose we introduce
local canonical coordinates $(p,q)$ around $f$ and $(\Pb,\Qb)$ around
$\fb$, so that $(\Pb(p,q,b),\Qb(p,q,b))=\chi_{_{ij,b}}(p,q)$, and
$v\Pb$ is the
vector field $v$ operating on $\Pb$ as a function on $B$. In these
coordinates,
$$\align \b(b,f)&=\<\a_F(\fb),J_*v\>=\Pb_mv(\Qb_m)\tag{\eqlab{beta}}\\
\Ham_\b&=\left[\frac{\p\Pb_m}{\p p_n}v(\Qb_m)
+\Pb_m\frac{\p v(\Qb_m)}{\p p_n}\right]\frac{\p\ }{\p q_n}
-\left[\frac{\p \Pb_m}{\p q_n}v(\Qb_m)
+\Pb_m\frac{\p v(\Qb_m}{\p q_n}\right]\frac{\p\ }{\p p_n}\endalign$$
Since $\chi_{_{ij,b}}$ preserves the canonical one--form $\a_F$, we have
$$\align \Pb_m\frac{\p\Qb_m}{\p p_n}=p_n,&\qquad
\Pb_m\frac{\p \Qb_m}{\p p_n}=0,\\
 \{\Pb_m,\Qb_n\}=\delta_{mn},&
\qquad \{\Pb_m,\Pb_n\}=\{\Qb_m,\Qb_n\}=0,\tag{\eqlab{caneq}}\endalign$$
where $\{,\}$ denotes the Poison bracket on $F$.
Using (\eqlab{caneq}), one can show that
$$\<\a_F(f),\Ham_\b\>=\Pb_mv(\Qb_m),\qquad 
\{\Ham_\b,\Pb_m\}=v(\Pb_m),\qquad\{\Ham_\b,\Qb_m\}=v(\Qb_m).
\tag{\eqlab{bracket}}$$
With (\eqlab{beta}) and (\eqlab{bracket}), we get
$$\align -i\Op(\b(b,f))\phi_\mu(\fb)&=\Ham_\b\phi_\mu(\fb)
+i\left(\<\a_F(f),\Ham_\b\>-\b\right)\phi_\mu=\Ham_\b\phi_\mu(\fb)\\
&=\frac{\p\phi_\mu(\fb)}{\p\Pb_m}\{\Ham_\b,\Pb_m\}+
\frac{\p\phi_\mu(\fb)}{\p\Qb_m}\{\Ham_\b,\Qb_m\}\\
&=\frac{\p\phi_\mu(\fb)}{\p\Pb_m}v(\Pb_m)
+\frac{\p\phi_\mu(\fb)}{\p\Qb_m}v(\Qb_m)=v(\phi_\mu(\fb))\\
&=v[\Chi_{\mu\sigma}]\phi_\sigma(f)\endalign$$
and the second term in (\eqlab{split}) becomes
$$\align
-i\ll\phi_\nu(\fb)\,|\,\Op(\b(b,f))\phi_mu(\fb)\gg
&=\ll\Chi_{\nu\tau}\phi_\tau(f)\,|\,v[\Chi_{\mu\sigma}]\phi_\sigma(f)\gg\\
&=\overline\Chi_{\nu\tau}v[\Chi_{\mu\tau}]=
v[\Chi_{\mu\tau}]\overline\Chi^t_{\tau\mu}\endalign$$
which in matrix notation is $v\Chi\Chi^{-1}$.
\qed\enddemo
\demo{Proof of proposition \thmlab{covariant}}
Let $\psi=\Psi_\mu\phi_\mu$ be a section on $\L$ covariant constant
along $v^\#$, let $w=\<\a^i,v\>$, then on $U_i\times F$ we have
$$\align
(v-\Ham_w)\Psi_\mu\phi_\mu&=i(\<\a^i,v\>
+\<\a_B,v\>-\<\a_F,\Ham_w\>)\Psi_\mu\phi_\mu\\
(v\Psi_\mu)\phi_\mu-\Psi_\mu(\Ham_w\phi_\mu)&=iw\Psi_\mu\phi_\mu
+i\<\a_B,v\>\Psi_\mu\phi_\mu-i\<\a_F,\Ham_w\>\Psi_\mu\phi_\mu\\
(v\Psi_\mu)\phi_\mu&=i\left(-i\Ham_w\phi_\mu-
\<\a_F,\Ham_w\>\phi_\mu+\<\a_B,v\>\phi_\mu+w\phi_\mu\right)\Psi_\mu\\
&=i(\<\a_B,v\>\phi_\mu+\Op(w)\phi_\mu)\Psi_\mu
\endalign$$
by (\eqlab{prequan}). Thus
$$\align
\ll\phi_\nu\,|\,(v\Psi_\mu)\phi_\mu\gg
&=i\ll\phi_\nu\,|\,\<\a_B,v\>\phi_\mu+\Op(w)\phi_\mu\gg
\Psi_\mu\\
v\Psi_\nu(b)&=\Psi_\mu(b)\left(i\<\a_B,v\>\delta_{\mu\nu}+
\left[A^i(b)v\right]_{\mu\nu}\right)\endalign$$
according to the definition (\eqlab{connection}).\qed
\vfil\pagebreak

\centerline{\bf References}
\vskip.4cm

\widestnumber\key{WW}

\ref \no\reflab{GLSW}
\by  M\. Gotay, R\. Lashof, J\. \'Sniatycki and A\. Weinstein
\pages 617(1983)
\vol 58
\jour Comment\. Math\. Helvetici\.
\endref

\ref \no\reflab{Sn}
\by J\. \'Sniatycki
\book Geometric quantization and quantum mechanics
\publ (Springer-Verlag, New York, 1980)
\endref

\ref \no\reflab{woodhouse}
\by N\.M\.J\. Woodhouse
\book Geometric Quantization
\publ (Clarendon Press, Oxford, second edition, 1992)
\endref

\ref \no\reflab{littlejohn}
\by R\.G\. Littlejohn and W\.G\. Flynn
\jour Phys\. Rev\. A
\vol 44
\pages 5239(1991)
\endref

\ref \no\reflab{BKS}
\by R\.J\. Blattner
\pages 87(1973)
\vol 26
\jour Proc\. Symp\. Pure Math\.
\endref

\ref \no\reflab{Robson}
\by  M\.A\. Robson
\pages 207
\yr 1996
\vol 19
\jour J\. Geom\. Phys\. 
\endref

\ref \no\reflab{MW}
\by J\. Marsden and A\. Weinstein
\pages 121-130
\yr 1974
\vol 5
\jour Rep\. Math\. Phys\.
\endref

\ref \no\reflab{Ki}
\by A\.A\. Kirillov
\book Elements of the theory of representation
\publ  (Springer-Verlag, Berlin, 1976)
\endref

\ref \no\reflab{Wu}
\by Y\. Wu
\paper Quantization of a particle in a background Yang-Mills field
\jour Jour\. Math\. Phys\.
\toappear
\paperinfo
e-Print Archive: quant-ph/9706040 
\endref
\end